\documentclass{article}
\usepackage{spconf,graphicx,hyperref}

\usepackage{algorithm}
\usepackage{algorithmic}
\usepackage{inconsolata}
\usepackage{amsmath,amssymb,amsfonts}
\usepackage{booktabs}
\usepackage{multirow}
\usepackage{multicol}
\usepackage{rotating}
\usepackage{array}     % Allows better text wrapping
\usepackage{xcolor}
\usepackage{colortbl}
\usepackage{tcolorbox}
\usepackage{subcaption}
\usepackage{enumitem}
\usepackage{balance}
\definecolor{Gray}{gray}{0.9}
\definecolor{orange}{rgb}{1.0, 0.647, 0.0}
\newcommand{\DefinedAI}{ContactCenter}

\definecolor{speech}{HTML}{2E68DB}
\definecolor{prompt}{HTML}{C73500}
\definecolor{transcript}{HTML}{666666}

\title{Reducing Prompt Sensitivity in LLM-based Speech Recognition Through Learnable Projection}
\name{
 \begin{tabular}{c}
 Sergio Burdisso$^{1,^\star}$, Esa\'u Villatoro-Tello$^{1,^\star}$,  Shashi Kumar$^{1,2}$, Srikanth Madikeri$^{4}$, Andr\'es Carofilis$^{1}$ \\
 Pradeep Rangappa$^{1}$, Manjunath K E$^{3}$, Kadri Hacioglu$^{3}$, Petr Motlicek$^{1,5}$, Andreas Stolcke$^{3}$
 \end{tabular}
}
\address{$^{1}$Idiap Research Institute \hspace{0.15cm}
$^{2}$EPFL \hspace{0.15cm}
$^{3}$Uniphore \hspace{0.15cm}
$^{4}$University of Zurich \hspace{0.15cm}
$^{5}$Brno University of Technology
}
\begin{document}
\ninept
\maketitle
\begin{abstract}
LLM-based automatic speech recognition (ASR), a well-established approach, connects speech foundation models to large language models (LLMs) through a speech-to-LLM projector, yielding promising results. A common design choice in these architectures is the use of a fixed, manually defined prompt during both training and inference. This setup not only enables applicability across a range of practical scenarios, but also helps maximize model performance.
However, the impact of prompt design remains underexplored.
This paper presents a comprehensive analysis of commonly used prompts across diverse datasets, showing that prompt choice significantly affects ASR performance and introduces instability, with no single prompt performing best across all cases. Inspired by the speech-to-LLM projector, we propose a prompt projector module, a simple, model-agnostic extension that learns to project prompt embeddings to more effective regions of the LLM input space, without modifying the underlying LLM-based ASR model.
Experiments on four datasets show that the addition of a prompt projector consistently improves performance, reduces variability, and outperforms the best manually selected prompts.
\end{abstract}
\begin{keywords}
LLM-based speech recognition, prompt sensitivity, speech-to-LLM projection, prompt projector, ASR robustness
\end{keywords}
\renewcommand*{\thefootnote}{\fnsymbol{footnote}}
\section{Introduction}\footnotetext{$^\star$ Corresponding authors: \{\textit{sergio.burdisso, esau.villatoro\}@idiap.ch}}
\renewcommand*{\thefootnote}{\arabic{footnote}}
\label{secc:Introduction}

Integrating speech capabilities into large language models (LLMs) is a key research area, enabling seamless voice interaction and enhancing multimodal understanding~\cite{goel2025audio,chu2024qwen2,ma2024embarrassingly,wang2023slm}. A conventional approach uses a cascaded architecture, first with an automatic speech recognition (ASR) system followed by an LLM to process the recognized text~\cite{huang2024multilingual, li2023prompting, ma2023can, yang2023generative}. However, this pipeline suffers from error propagation and prevent the LLM from having access to prosodic information that could be relevant for downstream tasks.

A promising alternative is LLM-based ASR, which directly connects a speech foundation model to an instruction-tuned LLM via a lightweight \textit{speech projector}~\cite{tangsalmonn, wu2023decoder, ma2024embarrassingly}. This projector maps speech-derived embeddings into the LLM's input space, enabling direct conditioning on the audio signal alongside textual prompts to perform ASR. While large audio language models (LALMs) are designed for general audio understanding with diverse prompts, LLM-based ASR systems focus specifically on transcription, relying on one \textit{fixed}, manually defined prompt to be used during both training and inference~\cite{ma2024embarrassingly, 10445874, kumar2024performance, yang2025bridging, fang2025low, sedlavcek2025approaching, wang2023slm, yang24f_interspeech, burdisso2025_text_only}.
This fixed-prompt setup ensures alignment between training objectives and inference behavior, making it well-suited for applications where high-accuracy transcription is the primary goal.

While prior work has explored different speech encoders and projectors~\cite{chu2024qwen2,das2024speechverse,10800077}, a critical component has been largely overlooked: the \textit{manually chosen prompt}. To the best of our knowledge, no prior work has systematically studied its impact on ASR performance.
In this paper, we take a first step toward understanding and improving prompt robustness in LLM-based ASR. We address two key research questions: (i) How important is the choice of the prompt for ASR performance? and (ii) Can we naturally extend a typical LLM-based ASR architecture to improve its robustness to prompt choice?

To answer the first question, our comprehensive evaluation of prompts from recent literature across multiple datasets reveals that prompt choice can drastically impact ASR output, with certain prompts yielding significantly better performance even under identical model and data conditions.

To address the second question, we propose a \textit{prompt projector} as a simple, yet effective, architectural extension. Inspired by the success of the speech projector, we hypothesize that a similar projection function can align prompt embeddings to a more effective region of the LLM input space. Our approach is distinct from soft-prompt learning, as it learns a common projection rather than individual embeddings.

Our main contributions are the following: (1) We provide the first systematic analysis of prompt sensitivity in LLM-based ASR systems; (2) we extend the original architecture by introducing the prompt projector; (3) we validate it across five evaluation sets, yielding consistently increased robustness and mitigating prompt-induced performance variance; (4) we release our source code for reproducibility.\footnote{\url{https://github.com/idiap/llm-asr-prompt}}

\begin{figure}
    \centering
    \includegraphics[width=0.9\linewidth]{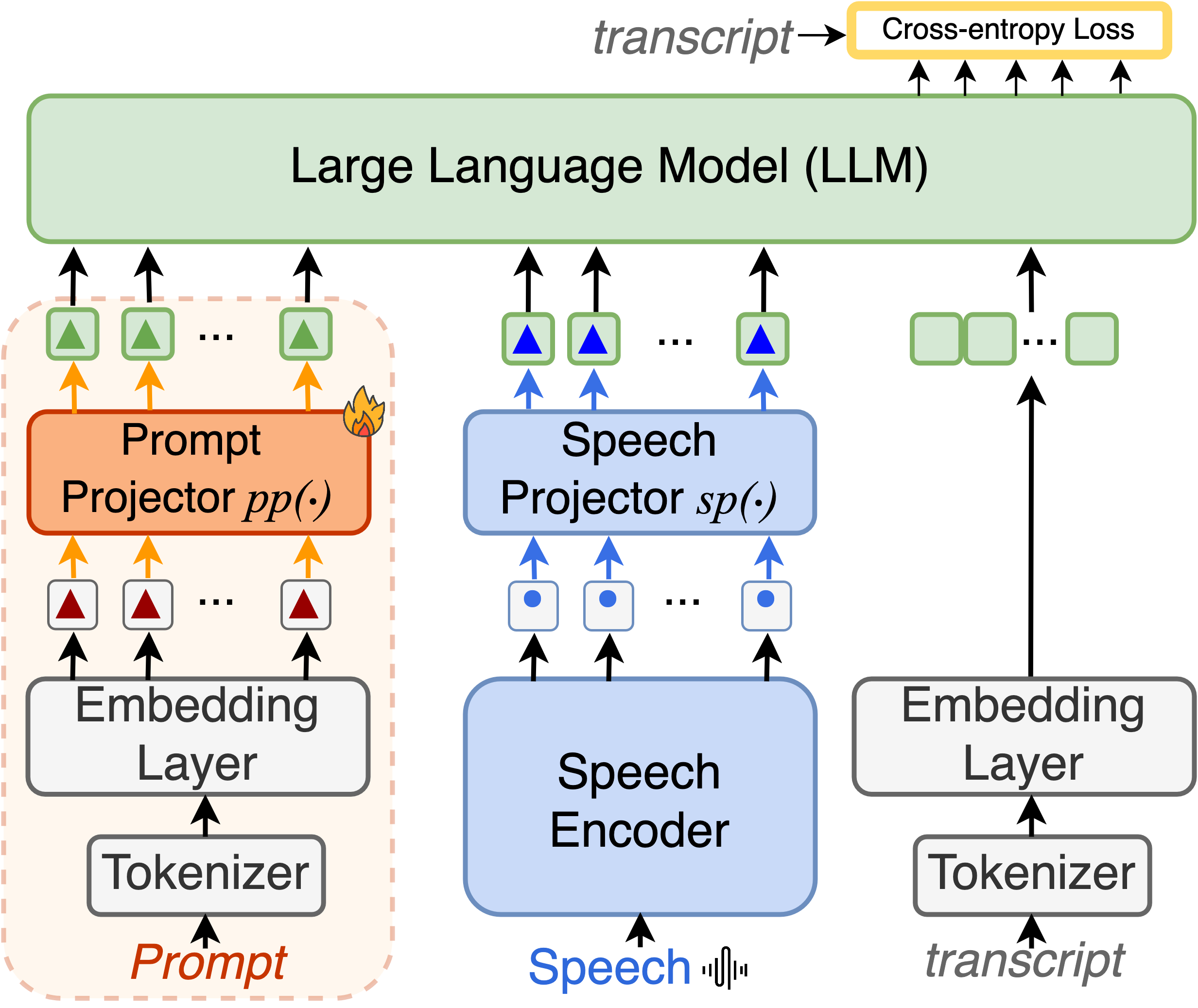}
    \caption{Typical LLM-based ASR system composed of a fixed prompt, a speech encoder, and an LLM connected by a speech projector, \textcolor{speech}{$sp(\cdot)$}.
    The proposed extension, highlighted in orange, introduces a learnable prompt projector, \textcolor{prompt}{$pp(\cdot)$}, into the original (frozen) architecture.
    The \textcolor{prompt}{$pp(\cdot)$} learns a common/single projection to transform all the original prompt embeddings (\textcolor{purple}{$\blacktriangle$}) into more effective ones (\textcolor{teal}{$\blacktriangle$}).}
    \label{fig:SLAM-v2}
\end{figure}

% \section{LALMs and LLM-based ASRs}

% Large Audio Language Models (LALMs) and LLM-based ASR represent two distinct approaches to integrating speech with large language models. While both leverage LLMs, they differ in scope and prompting strategies.

% LALMs are designed for general-purpose audio understanding, such as audio captioning, event detection, and spoken language reasoning. They are typically trained via multimodal instruction tuning on datasets consisting of audio-instruction-response triples~\cite{zhang-etal-2023-speechgpt, goel2025audio, das2024speechverse, chu2024qwen2}, allowing them to handle diverse prompts and multitask scenarios.

% In contrast, LLM-based ASR focuses on transcription, using a fixed, manually defined prompt during both training and inference~\cite{ma2024embarrassingly, 10445874, kumar2024performance, yang2025bridging, fang2025low, sedlavcek2025approaching, wang2023slm, yang24f_interspeech}. This fixed-prompt setup ensures alignment between training objectives and inference behavior, making it well-suited for applications where high-accuracy transcription is the primary goal. Consequently, our research is framed around this fixed-prompt formulation.

% Despite its narrower focus, LLM-based ASR enables the combination of strong pre-trained speech encoders with powerful LLMs via the learnable projection, benefiting from advances in both fields. This modular design supports scalable, high-performance transcription without costly instruction tuning.

\section{Methodology}
\subsection{ 
Base Model}
\label{sec:slam}

As our base model, we adopt the recently proposed SLAM-ASR architecture for LLM-based ASR~\cite{ma2024embarrassingly}.%\footnote{\url{https://github.com/X-LANCE/SLAM-LLM/blob/main/examples/asr_librispeech/README.md}}
This model showed that a simple speech projection module is sufficient to achieve competitive performance, outperforming more complex LLM-based ASR systems. Its strong results and minimal design make it an ideal foundation for our experiments.

Specifically, the speech projector, \textcolor{speech}{$sp(\cdot)$}, is defined as:
\begin{equation}
\label{eq:projector}
    \mathbf{e}_i = sp(\mathbf{z}_i) = \text{ReLU}(\mathbf{z}_iW_1 + b_1)W_2 + b_2,
\end{equation}
where $\mathbf{z}_i$ denotes the $i$-th downsampled audio feature, obtained by concatenating $k$ consecutive encoded frames along the temporal dimension. The resulting projection $\mathbf{e}_i$ matches the dimensionality of the LLM input embeddings.

In the original work~\cite{ma2024embarrassingly}, an input audio consisting of $n$ downsampled features, $\mathbf{z}_1, \dots, \mathbf{z}_n$, is fed to the LLM using the following prompt, which we refer to as the \textit{``base''} prompt:\footnote{The original paper places speech embeddings after \texttt{"USER:"}, but \href{https://github.com/X-LANCE/SLAM-LLM/tree/main/examples/asr_librispeech}{the released code} prepends them. Our prompt here reflects their implementation.}

\begin{description}[noitemsep]
\begin{small}
    \item \texttt{\textbf{\textcolor{speech}{\{speech\}}}\textcolor{prompt}{<s>USER: Transcribe speech to text.}}
    \item \texttt{\textcolor{prompt}{ASSISTANT:} \textit{\textcolor{transcript}{\{transcript\}}\textcolor{prompt}{</s>}}}
\end{small}
\end{description}
where \texttt{\textit{\textcolor{transcript}{\{transcript\}}}} is the transcription corresponding to the audio, which is either provided during training or forced to be generated by the LLM at inference time.
Here, \textcolor{speech}{\texttt{\textbf{\{speech\}}}} is replaced by the sequence of $n$ projected speech embeddings, \textcolor{speech}{$sp(\mathbf{z}_1), \dots, sp(\mathbf{z}_n)$}, when the prompt is provided to the LLM.

We use the same speech encoder and LLM as in the original work, as this configuration has been shown to outperform models using larger and more recent LLMs~\cite{ma2024embarrassingly, yang24f_interspeech, kumar2024performance}, offering a good trade-off between ASR performance and computational efficiency:

% \begin{itemize}
%     \item \textbf{WavLM-large}\footnote{\url{https://huggingface.co/microsoft/wavlm-large}}: speech encoder trained on 94k hours of unlabeled data using self-supervision~\cite{chen2022wavlm,kahn2020libri,wang-etal-2021-voxpopuli,chen2021gigaspeech}.
%     \item \textbf{Vicuna-7B}\footnote{\url{https://huggingface.co/lmsys/vicuna-7b-v1.5}}: LLM fine-tuned from LLama with SFT and optionally RLHF~\cite{chiang2023vicuna}.\footnote{We tested Llama 3 8B, but Vicuna 7B still outperformed it on 3 of the 5 evaluation sets, so we stick to the original paper’s setup. Details on \href{https://github.com/idiap/llm-asr-prompt}{our Github}.}
% \end{itemize}

%%\noindent $\bullet$ \textbf{WavLM-large}\footnote{\url{https://huggingface.co/microsoft/wavlm-large}}: speech encoder trained on large-scale unlabeled speech data using self-supervision techniques~\cite{chen2022wavlm}.
%%More precisely, the model was trained on 94,000 hours of data, including LibriLight~\cite{kahn2020libri}, VoxPopuli~\cite{wang-etal-2021-voxpopuli}, and GigaSpeech~\cite{chen2021gigaspeech}, making it a robust speech foundation model.
\noindent $\bullet$ \textbf{WavLM-large}\footnote{\url{https://huggingface.co/microsoft/wavlm-large}}: speech encoder trained on 94k hours of unlabeled data using self-supervision~\cite{chen2022wavlm,kahn2020libri,wang-etal-2021-voxpopuli,chen2021gigaspeech}.

%%\noindent $\bullet$ \textbf{Vicuna-7B}\footnote{\url{https://huggingface.co/lmsys/vicuna-7b-v1.5}}: LLM fine-tuned from LLama using supervised fine-tuning (SFT) on user-shared conversations and optionally enhanced with reinforcement learning from human feedback (RLHF) to improve response quality and alignment with human preferences~\cite{chiang2023vicuna}.\footnote{We tested Llama 3 8B, but Vicuna 7B still outperformed it on 3 of the 5 evaluation sets, so we stick to the original paper’s setup.}
\noindent $\bullet$ \textbf{Vicuna-7B}\footnote{\url{https://huggingface.co/lmsys/vicuna-7b-v1.5}}: LLM fine-tuned from LLama with SFT and optionally RLHF~\cite{chiang2023vicuna}.\footnote{We tested Llama 3 8B, but Vicuna 7B still outperformed it on 3 of the 5 evaluation sets, so we stick to the original paper’s setup.}

%In summary, throughout all experiments in this paper, we consistently use the \textit{WavLM-large + Vicuna-7B} combination. As in the original work~\cite{ma2024embarrassingly}, and unless stated otherwise, both WavLM-large and Vicuna-7B are kept frozen during the training of the projector in Equation~\ref{eq:projector}. The original work also presents a comprehensive benchmark across various speech encoders and LLMs, identifying this configuration as optimal in terms of WER performance, without requiring fine-tuning of the speech encoder.
%We refer to the model described here as the \textit{``vanilla''} model throughout this paper.
Throughout all experiments, we use the \textit{WavLM-large + Vicuna-7B} combination.
This configuration was shown to be optimal in WER without fine-tuning the speech encoder~\cite{ma2024embarrassingly}.
We refer to this model as the \textit{``vanilla''} model and, as in the original work, it is trained by only learning the projector  (Equation~\ref{eq:projector}) while keeping the speech encoder and the LLM frozen.

\subsection{The Prompt Projector: \textcolor{prompt}{$pp(\cdot)$}}
\label{subsecc:promptor}

To reduce variability caused by prompt choice (Section \ref{sec:experimentation}), we introduce \textcolor{prompt}{$pp(\cdot)$}, a \textit{prompt projector} that projects original prompt embeddings into a more effective region of the LLM input space (Figure~\ref{fig:SLAM-v2}).

The prompt projector module is a simple drop-in extension to an existing LLM-based ASR system: after training the base model—or using a pretrained one—we freeze all components and train only the new projector module. This ensures the training only focuses on learning how to project the prompt embeddings and avoids instability.\footnote{In complementary experiments, we found out that unfreezing the underlying models consistently led to unstable training and degraded performance across all datasets. Detailed results can be found on \href{https://github.com/idiap/llm-asr-prompt}{our Github}.}

Formally, given prompt embeddings \(\textcolor{prompt}{\mathbf{x}_1, \dots, \mathbf{x}_n}\), they are projected via \(\textcolor{prompt}{pp(\cdot)}\) and the new sequence \(\textcolor{prompt}{pp(\mathbf{x}_1), \dots, pp(\mathbf{x}_n)}\) is passed to the LLM instead. The projector shares the same architecture as the speech projector \(\textcolor{speech}{sp(\cdot)}\) (Equation~\ref{eq:projector}), differing only in input dimensionality, since it operates on LLM embeddings rather than speech features.

This simple, architecture-consistent design improves robustness to prompt choice without modifying the original system or introducing prompt-engineering parameters or tokens.

\begingroup
\renewcommand{\arraystretch}{1.30}
\begin{table}[!t]
    \centering
    \scriptsize
        \caption{The set of 10 prompts considered for experimentation. The \textcolor{speech}{\texttt{\textcolor{speech}{\textbf{\{speech\}}}}} indicates where the (projected) speech embeddings are located within the prompt. \texttt{\textit{\textcolor{transcript}{\{transcript\}}}}s are omitted for simplicity.}
    \label{tab:prompts}
    \begin{tabular}{c|p{6.8cm}}
        \toprule
        \textbf{No.} & \textbf{Prompt Template} \\
        \hline
        \textit{empty} & \texttt{\textcolor{speech}{\textbf{\{speech\}}}} \\
        \hline
        \textit{base} & \texttt{\textcolor{speech}{\textbf{\{speech\}}}\textcolor{prompt}{<s>USER: Transcribe speech to text.\textbackslash n ASSISTANT:}} \\
        \hline
        1 & \texttt{\textcolor{prompt}{<s>USER: Transcribe speech to text.} \textcolor{speech}{\textbf{\{speech\}}}\textcolor{prompt}{\textbackslash n ASSISTANT:}} \\
        \hline
        2 & \texttt{\textcolor{prompt}{<s>USER: Transcribe speech to text. Speech:} \textcolor{speech}{\textbf{\{speech\}}}\textcolor{prompt}{.\textbackslash n ASSISTANT:}} \\
        \hline
        3 & \texttt{\textcolor{prompt}{<s>USER: Transcribe the following speech to text:} \textcolor{speech}{\textbf{\{speech\}}}\textcolor{prompt}{.\textbackslash n ASSISTANT:}} \\
        \hline
        4 & \texttt{\textcolor{prompt}{<s>USER: Transcribe accurately speech to text. English speech:} \textcolor{speech}{\textbf{\{speech\}}}\textcolor{prompt}{.\textbackslash n ASSISTANT:}} \\
        \hline
        5 & \texttt{\textcolor{prompt}{<s>USER: Audio:} \textcolor{speech}{\textbf{\{speech\}}}\textcolor{prompt}{.\textbackslash n Transcribe the preceding audio.\textbackslash n ASSISTANT:}} \\
        \hline
        6 & \texttt{\textcolor{prompt}{<s>USER: Audio:} \textcolor{speech}{\textbf{\{speech\}}}\textcolor{prompt}{.\textbackslash n What is being said in the preceding audio?\textbackslash n ASSISTANT:}} \\
        \hline
        7 & \texttt{\textcolor{prompt}{<s>USER: Transcribe the following audio:} \textcolor{speech}{\textbf{\{speech\}}}\textcolor{prompt}{.\textbackslash n ASSISTANT:}} \\
        \hline
        8 & \texttt{\textcolor{prompt}{<s>USER: What is being said in the following audio? Audio:} \textcolor{speech}{\textbf{\{speech\}}}\textcolor{prompt}{.\textbackslash n ASSISTANT:}} \\
        \bottomrule
    \end{tabular}
\end{table}
\endgroup

\begin{table*}[ht!]
    \centering
    % \footnotesize
    \scriptsize
        \caption{WER (\%) comparison of the \textit{vanilla} model with and without the prompt projection (+\textcolor{prompt}{$pp(\cdot)$}) across different prompts. Relative improvement ($\Delta\%$) after applying \textcolor{prompt}{$pp(\cdot)$} is also reported. \textbf{Bold} numbers indicate the best performance in each column, while \underline{\textit{underlined}} values highlight the largest relative improvement.}

        %\caption{WER (in \%) comparison of the \textit{vanilla} model \textit{vs.} before and after applying the prompt projection (+\textcolor{prompt}{$pp(\cdot)$}) using different prompts. Relative improvement after applying \textcolor{prompt}{$pp(\cdot)$} is also reported ($\Delta\%$). Numbers in \textbf{bold} indicate the best performance for each column, while \underline{\textit{underlined}} values highlight the maximum relative improvement.}
        \label{tab:prompt_template_results}
    \renewcommand{\arraystretch}{1.4}
     \setlength{\tabcolsep}{2.8pt}
    \begin{tabular}{c|ccc|ccc|ccc|ccc|ccc}
        \toprule
        %\multirow{4}{*}{\textbf{Prompt}} & \multicolumn{15}{c}{\textbf{Word Error Rate (WER\% $\downarrow$) \& Relative Improvement values ($\Delta\% \uparrow$)}} \\
        \cmidrule(lr){2-16}
        \multirow{3}{*}{\textbf{Prompt}} &  \multicolumn{3}{c}{\textbf{ContactCenter}} & \multicolumn{3}{c}{\textbf{CallHome}} & \multicolumn{3}{c}{\textbf{AMI}} & \multicolumn{3}{c}{\textbf{LibriSpeech-Clean}} & \multicolumn{3}{c}{\textbf{LibriSpeech-Other}} \\
        \cmidrule(lr){2-4}\cmidrule(lr){5-7}\cmidrule(lr){8-10}\cmidrule(lr){11-13}\cmidrule(lr){14-16}
        & \textit{vanilla} & +\textcolor{prompt}{$pp(\cdot)$} & $\Delta\%$ & \textit{vanilla} & +\textcolor{prompt}{$pp(\cdot)$} & $\Delta\%$ & \textit{vanilla} & +\textcolor{prompt}{$pp(\cdot)$} & $\Delta\%$ & \textit{vanilla} & +\textcolor{prompt}{$pp(\cdot)$} & $\Delta\%$ & \textit{vanilla} & +\textcolor{prompt}{$pp(\cdot)$} & $\Delta\%$\\
        \midrule 
        \textit{empty} & 12.75& - & - & 27.00& - & - & 13.88& - & - & 2.84 &- & - & 5.40 &- & -\\
        % \textit{base} & 13.00 & 11.53 & \textit{(11.3)} & 29.26 & 27.15 & \textit{(7.2)} & 13.86 & 13.39 & \textit{(3.4)} & 3.09 & 2.34 & \textit{\underline{(24.3)}} & 5.85 & 4.98 & \textit{\underline{(14.9)}}\\
        \textit{base} & 13.00 & 11.23 & \textit{(11.3)} & 29.26 & 26.52 & \textit{(7.2)} & 13.86 & 13.42 & \textit{(3.4)} & 3.09 & 2.34 & \textit{\underline{(24.3)}} & 5.85 & 4.98 & \textit{\underline{(14.9)}}\\
        \textbf{1} & 11.91 & 11.58 & \textit{(2.8)} & \textbf{25.26} & 24.84 & \textit{(1.7)} & 13.72& 12.96 & \textit{(5.5)} & 2.88 &2.39 & \textit{(17.0)} & 5.59 &4.89 & \textit{(12.5)} \\
        \textbf{2} & 12.27 & 11.31 & \textit{(7.8)} & 27.08 &\textbf{24.73} & \textit{\underline{(8.7)}} & \textbf{13.36}& 12.78 & \textit{(4.3)} & 2.89 &2.31 & \textit{(20.1)} & 5.71 &4.84 & \textit{(15.2)} \\
        \textbf{3} & \textbf{11.81} & 11.25 & \textit{(4.7)} & 25.83 & 25.90 & \textit{(-0.3)} & 13.50 & 13.26 & \textit{(1.8)} & \textbf{2.72} &2.31 & \textit{(15.1)} &\textbf{ 5.30} &4.92 & \textit{(7.2)} \\
        \textbf{4} & 12.68 &12.43 & \textit{(2.0)} & 27.95 &25.94 & \textit{(7.2)} & 13.83& 12.80 & \textit{\underline{(7.4)}} & 2.75 &\textbf{2.28} & \textit{(17.1)} & 5.38 &\textbf{4.79} & \textit{(11.0)} \\
        \textbf{5} & 12.71 & \textbf{11.23} & \textit{\underline{(11.6)}} & 25.77 &25.62 & \textit{(0.6)} & 13.54& 13.18 & \textit{(2.7)} & 2.80 &2.29 & \textit{(18.2)} & 5.42 &5.15 & \textit{(5.0)} \\
        \textbf{6} & 12.44 &11.73 & \textit{(5.7)} & 26.17 &25.57 & \textit{(2.3)} & 13.37& 12.77 & \textit{(4.5)} & 2.80 &2.36 & \textit{(15.7)} & 5.47 &5.04 & \textit{(7.9)} \\
        \textbf{7} & 12.30 &11.48 & \textit{(6.7)} & 26.69 &25.60 & \textit{(4.1)} & 13.49& 12.91 & \textit{(4.3)} & 2.95 &2.31 & \textit{(21.7)} & 5.37 &5.06 & \textit{(5.8)} \\
        \textbf{8} & 12.00 &11.44 & \textit{(4.7)} & 25.56 &24.93 & \textit{(2.5)} & 13.42& \textbf{12.74} & \textit{(5.1)} & 2.91 &2.61 & \textit{(10.3)} & 5.54 &5.14 & \textit{(7.2)} \\
        \bottomrule
    \end{tabular}
\end{table*}

\begin{figure}[t!]
    \centering
    \includegraphics[width=1\linewidth]{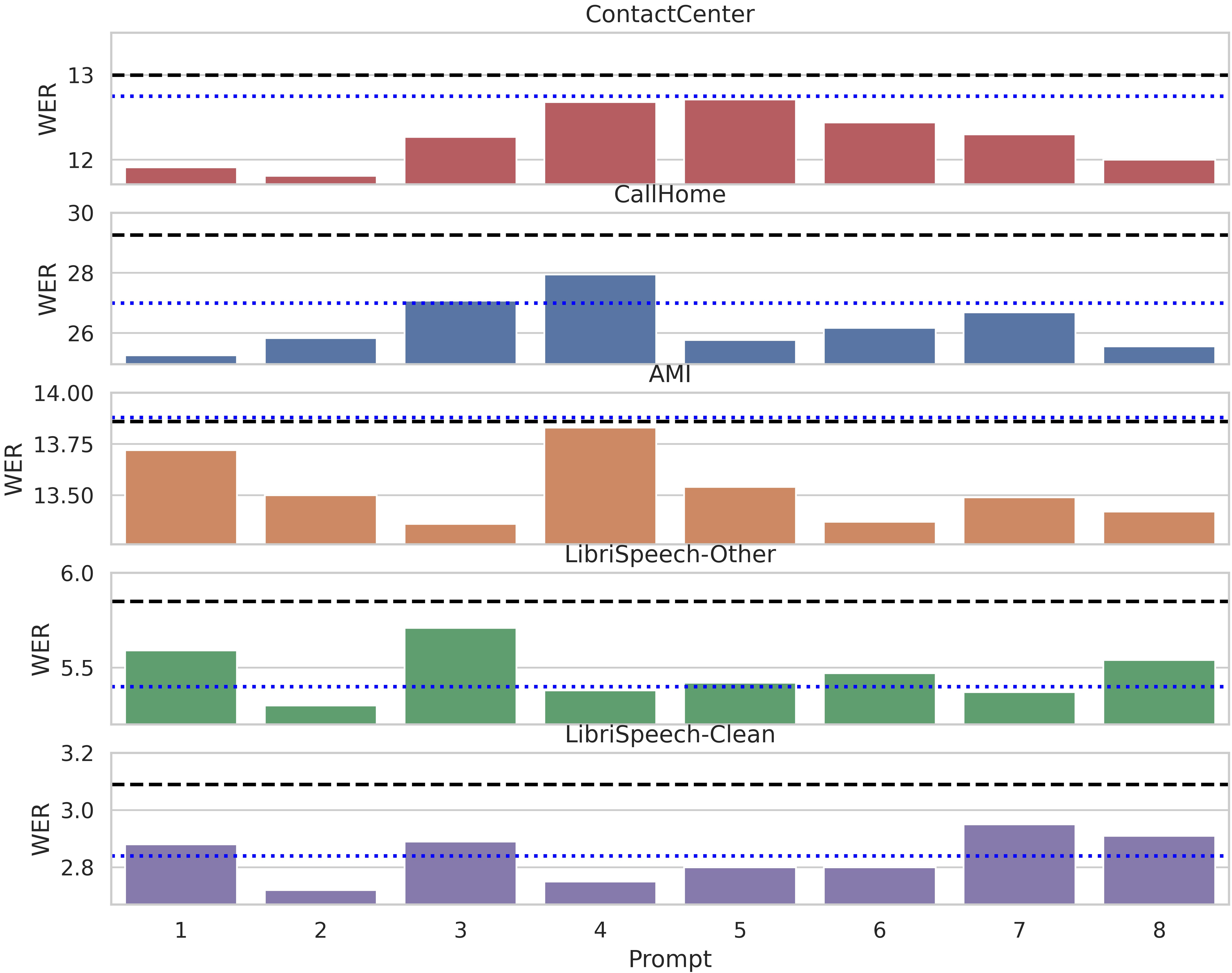}
    \caption{ASR performance (WER (in \%)) across datasets with different prompts. The black dashed line represents the \textit{base} prompt, while the blue dotted line corresponds to the \textit{empty} prompt.}
    \label{fig:prompt_impact}
\end{figure}

\subsection{Training \& Inference Implementation Details}
\label{sub:implementation}

%To ensure that our results are reproducible and comparable, in all our experiments and for each dataset, both speech and prompt projectors were trained for 5 epochs\footnote{Due to limited computational resources, the LibriSpeech (1K hours) experiments were trained for only one epoch.} on the same training set, with early stopping based on cross-entropy loss on the dev set, using AdamW~\cite{loshchilov2018decoupled} with a learning rate $\gamma=10^{-4}$ and a batch size of 4.

%The speech encoder produces output at 50 Hz and the downsampling rate is set to $k = 5$, as in the original work~\cite{ma2024embarrassingly}, leading to the downsampled audio features \textcolor{speech}{$\mathbf{z}_i$} having a rate of 10 per second ---\textit{i.e.} \textcolor{speech}{\texttt{\textbf{\{speech\}}}} will have 10 speech embeddings \textcolor{speech}{$sp(\mathbf{z}_i)$} per second of audio. The prompt and speech projectors hidden layer dimension is set to 2048 and beam search with a beam size of 4 is used for decoding.
%Additionally, for the experiments involving LLM fine-tuning using LoRA adapters, the rank is set to 8 and alpha to 32.

All experimental settings follow the original work~\cite{ma2024embarrassingly}. Specifically, the speech encoder generates the output at 50\,Hz with a downsampling rate of $k = 5$, producing downsampled audio features \textcolor{speech}{$\mathbf{z}_i$} at the 100\,ms rate — i.e., \textcolor{speech}{\texttt{\textbf{\{speech\}}}} contains 10 embeddings \textcolor{speech}{$sp(\mathbf{z}_i)$} per second of audio. Both speech and prompt projectors have a hidden layer dimension of 2048, and decoding uses beam search with size 4. Training uses AdamW~\cite{loshchilov2018decoupled} with learning rate $\gamma=10^{-4}$, batch size 4, and early stopping based on cross-entropy loss on the dev set.  

All computations were performed in \textit{bfloat16} format. Overall, our experiments required over 150 train-to-evaluation trials across various settings (10 prompts, 4 datasets, +/- \textcolor{prompt}{$pp(\cdot)$}, +/- LoRA, freezing/unfreezing) on a single NVIDIA H100 (80 GB VRAM). For our experiments, we trained all models for 5 epochs,%
\footnote{To limit computation given the large number of experiments, experiments using LibriSpeech (about 1k hours) trained for only one epoch, each taking about one day.}
and, in experiments involving LLM fine-tuning, we used LoRA adapters with rank 8 and $\alpha =32$.

%In addition, we trained the models for 5 epochs\footnote{Due to limited computational time, LibriSpeech (1K hours) experiments were trained for only one epoch} — and, in experiments involving LLM fine-tuning, the use of LoRA adapters with rank 8 and alpha 32.

% As main evaluation metric, we use word error rate (WER), a well established metric that evaluate the accuracy of a ASR system by quantifying how many errors (deletions, insertions and substitutions) the system makes when transcribing spoken audio into text. 
% To prevent artificially inflated WER caused by hallucinations in short segments (\textit{e.g.}, for instance, a single-word reference is decoded as the same word repeated up to the maximum length), we apply a simple post-processing normalization step to all models and results reported in the paper.
% Specifically, we calculate the word ratio ($wr=length(hyp)/voc(hyp)$, where $voc(\cdot)$ returns the number of unique words within a hypothesis generated by the LLM. If a hypothesis has a high $wr$ ($>7$), it is replaced with a single \texttt{unk} token.
% This ensures that the WER in such cases is correctly set to 100.
%Finally, all the model computations were performed using \textit{bfloat16} floating-point format and, overall, the experiments reported in this work required over 150 train $\rightarrow$ evaluation trials across various settings (10 prompts, 4 datasets, +/- \textcolor{prompt}{$pp(\cdot)$}, +/- LoRA, freezing/unfreezing) on single NVIDIA H100 (80Gb of VRAM) .%, each taking from 5 hours up to 1.5 days depending on the dataset and configuration.

\subsection{Datasets}
\label{secc:Datasets}

We used four datasets spanning diverse speaking styles (read, spontaneous, and conversational), domains, and recording conditions:
% \begin{itemize}
%     \item \textbf{LibriSpeech (LS)}: A large-scale corpus of 1000 hours of read English speech from audiobooks, sampled at 16 kHz \cite{panayotov2015librispeech}. Our models were trained on the 960h partition and evaluated on the \texttt{test-clean} (LS-C) and \texttt{test-other} (LS-O) subsets.
%     \item \textbf{CallHome (CH)}: This dataset contains 12.5h of spontaneous telephone conversations, which presents challenges for ASR due to its conversational style and numerous short segments \cite{callhome}.
%     \item \textbf{AMI}: A multi-modal dataset with 100 hours of meeting recordings. We used the individual head-mounted microphone (IHM) portion, with approximately 80 hours for training and 8.5 hours for evaluation \cite{ami}.
%     \item \textbf{\DefinedAI\ (CC)}: A proprietary dataset of 48h of contact center conversations from health and finance domains. The 8 kHz audio was upsampled to 16 kHz for compatibility with our baseline model. This dataset is particularly challenging due to its domain-specific nature.
% \end{itemize}
\noindent$\bullet$ \textbf{LibriSpeech (LS)}: A 1000\,h corpus of read English audiobooks~\cite{panayotov2015librispeech}. Models were trained on the 960\,h split and evaluated on the 5.4\,h \texttt{test-clean} (LS-C) and 5.1\,h \texttt{test-other} (LS-O) sets, with their respective 5.4\,h \texttt{dev-clean} and 5.3\,h \texttt{dev-other} sets.
\noindent$\bullet$ \textbf{CallHome (CH)}: 17.5\,h of spontaneous telephone conversations~\cite{callhome}, split into 13\,h training, 3\,h dev, and 1.5\,h test sets. Its conversational style and short utterances make it challenging for ASR.
\noindent$\bullet$ \textbf{AMI}: A 100\,h multi-modal meeting corpus~\cite{ami}. We used the individual head-mounted microphone (IHM) recordings, comprising 80\,h training, 8.5\,h test, and 8.8\,h dev sets.
\noindent$\bullet$ \textbf{\DefinedAI\ (CC)}: A 48\,h proprietary corpus of contact center conversations in health and finance. It is split into 30\,h training, 4\,h dev, and 6\,h test sets, and is difficult due to its domain specificity.

\section{Experimentation}
\label{sec:experimentation}

\subsection{Assessing the Impact of Manual Prompts}

We first examine the effect of fixed prompts on the word error rate (WER\%) of our \textit{vanilla} LLM-based ASR model (Section~\ref{sec:slam}). For each prompt in Table~\ref{tab:prompts}, we train and evaluate an independent \textit{vanilla} model per dataset, isolating the impact of prompt choice. In total, nine prompts beyond the \textit{base} prompt are evaluated, along with a \textit{empty} prompt containing only speech embeddings.

Prompts 1–4 are variations of the \textit{base} prompt with minor wording changes or different placement of the speech embeddings (\textcolor{speech}{\texttt{\textbf{\{speech\}}}}). Prompts 5–8 were adapted from prior works, such as SpeechVerse~\cite{das2024speechverse} and SpeechLLM~\cite{Rajaa_SpeechLLM_Multi-Modal_LLM}.

Results are reported in Table~\ref{tab:prompt_template_results} (\textit{vanilla} column) and visualized in Figure~\ref{fig:prompt_impact}. Even subtle changes, such as the difference between the \textit{base} prompt and prompt 1, yield noticeable improvements: relative WER reductions of 13.6\% (CallHome), 8.3\% (ContactCenter), 6.7\% (LibriSpeech-Clean), and 4.4\% (LibriSpeech-Other).  

Overall, prompts 1–8 outperform the \textit{base} prompt across all the datasets, suggesting the original \textit{base} prompt is suboptimal and may limit transcription quality. Figure~\ref{fig:prompt_impact} also reveals inconsistent performance: some prompts excel on certain datasets but underperform on others. For instance, prompt 1 performs well on CC and CH but poorly on AMI and LS. In some cases,  prompts may even degrade performance relative to having no prompt at all (\textit{e.g.}, \textit{base} or prompt 4 on CH).

Notably, this demonstrates that an LLM-based ASR model can operate effectively using only speech embeddings. The speech projector, even in its simple form (Equation~\ref{eq:projector}), can \textit{implicitly encode prompt-like information while mapping speech features to the LLM input space}.  
We recommend including a \textit{no prompt} baseline in future LLM-ASR research. It provides a fast, low-cost diagnostic to detect poorly performing prompts—like the \textit{base} prompt—early in development.

These results highlight the high variability of manual prompts: no single prompt performs optimally across all datasets, and even minor changes in wording or embedding placement can lead to large differences in WER.

\subsection{Incorporating the Prompt Projector}
\label{sub:exp-promptor}

The results from the previous section highlight the need to reduce sensitivity to prompt choice. We now evaluate whether the \textit{prompt projector} (Section~\ref{subsecc:promptor}) can learn a projection, \textcolor{prompt}{$pp(\cdot)$}, that maps prompts into more effective regions of the LLM input space. Following the prior setup, we train and evaluate independent models for each prompt and dataset, now including \textcolor{prompt}{$pp(\cdot)$}. The projector is trained on frozen \textit{vanilla} model checkpoints reported in previous section.\footnote{Unfreezing the underlying model degrades performance, as mentioned in Section~\ref{subsecc:promptor}. A detailed analysis is available in the appendix on \href{https://github.com/idiap/llm-asr-prompt}{our GitHub}.} Results are reported in Table~\ref{tab:prompt_template_results} under the ``+\textcolor{prompt}{$pp(\cdot)$}'' column. 

Table~\ref{tab:main_results} summarizes key results, comparing the \textit{vanilla} model to its variant with the \textit{prompt projector} for both the base and best prompts per dataset, and contextualize the findings relative to other recent work. It also includes \textit{empty} and \textit{base} prompts, along with a ``best'' row showing the top-performing manual prompt per dataset. ``\textit{+LoRA}'' results report performance when models are fine-tuned with LoRA adapters.

Comparing \textit{base} and \textit{best} under \textit{vanilla} highlights the substantial variability due to prompt choice (e.g., 13.00 vs.\ 11.81 WER\% on CC, 29.26 vs.\ 25.26 on CH). Optimal prompts (\textit{best} under \textcolor{prompt}{$pp(\cdot)$}) can approach the best published results, whereas suboptimal prompts may underperform the \textit{empty} prompt (e.g., \textit{base} vs. \textit{empty} on CH and LS). Prompt choice also affects LLM fine-tuning (\textit{base+LoRA} vs.\ \textit{best+LoRA}), creating a potential bottleneck (e.g., 28.18 vs.\ 24.74 WER\% on CH).

\begin{figure}[t!]
    \centering
    % First row
    \begin{subfigure}{0.1\textwidth}
        \centering
        \includegraphics[height=2.75cm]{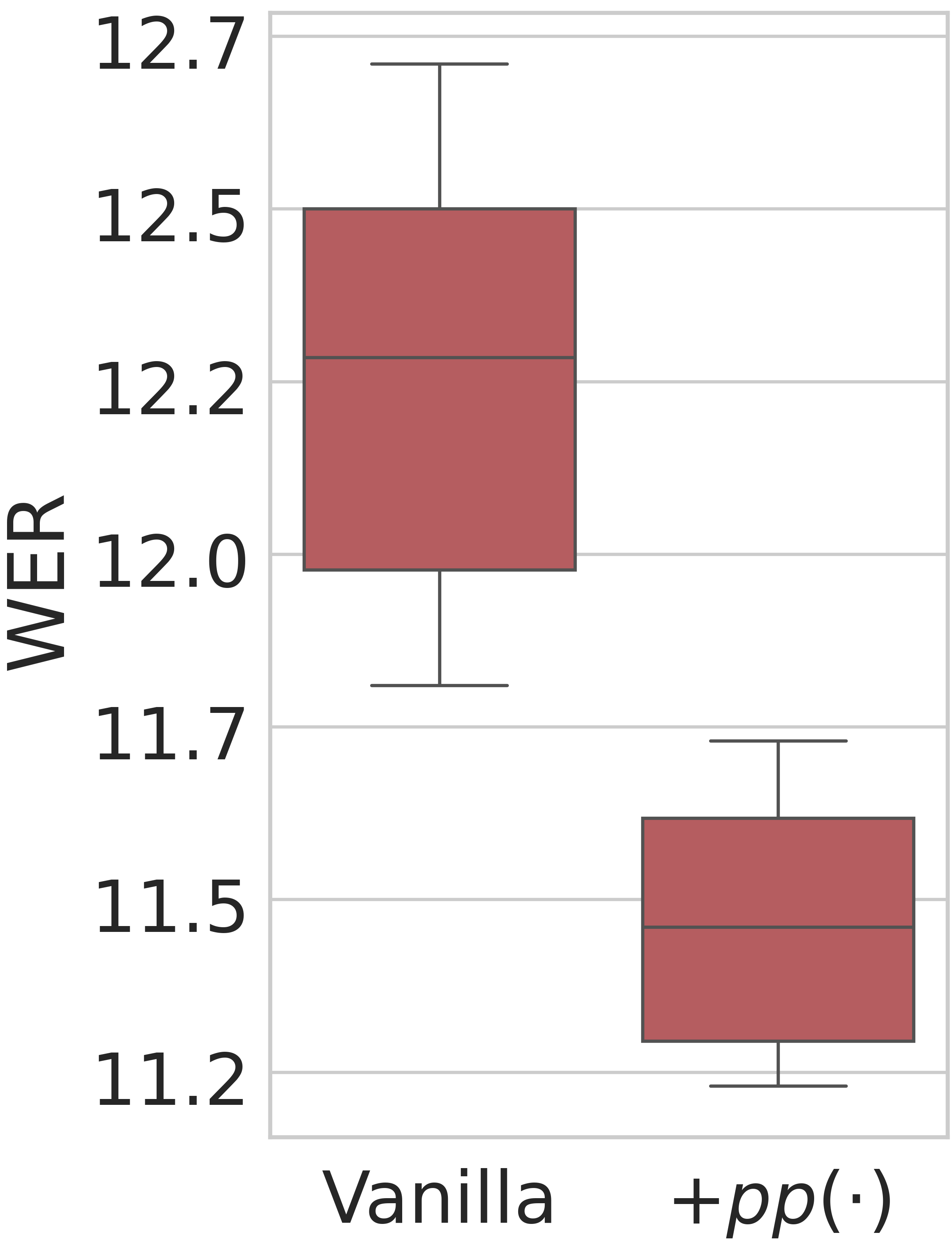}
        \caption{CC}
        \label{fig:subfig1}
    \end{subfigure}
    \hfill
    \begin{subfigure}{0.1\textwidth}
        \centering
        \includegraphics[height=2.75cm]{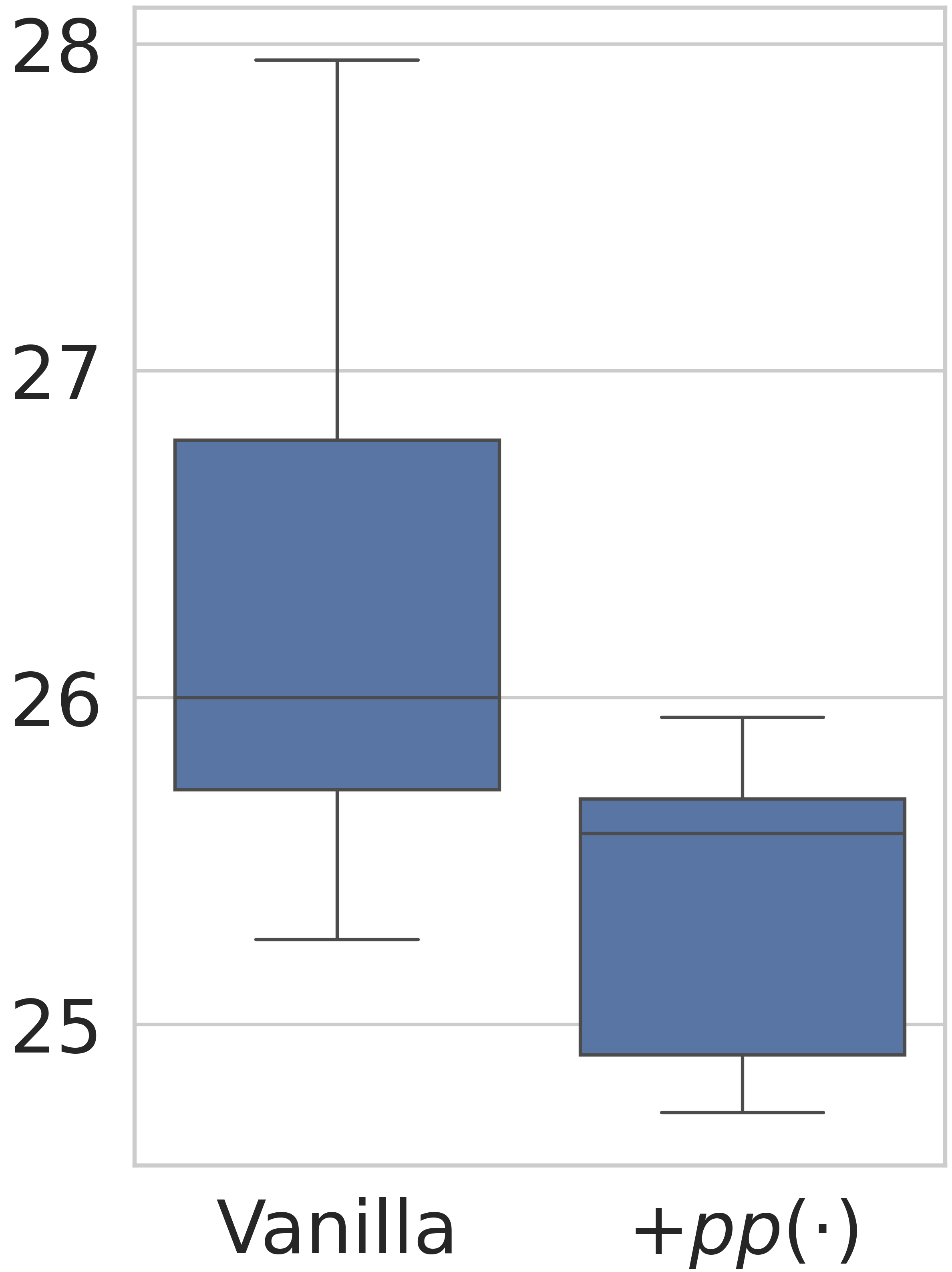}
        \caption{CH}
        \label{fig:subfig2}
    \end{subfigure}
    \hfill
    \begin{subfigure}{0.1\textwidth}
        \centering
        \includegraphics[height=2.75cm]{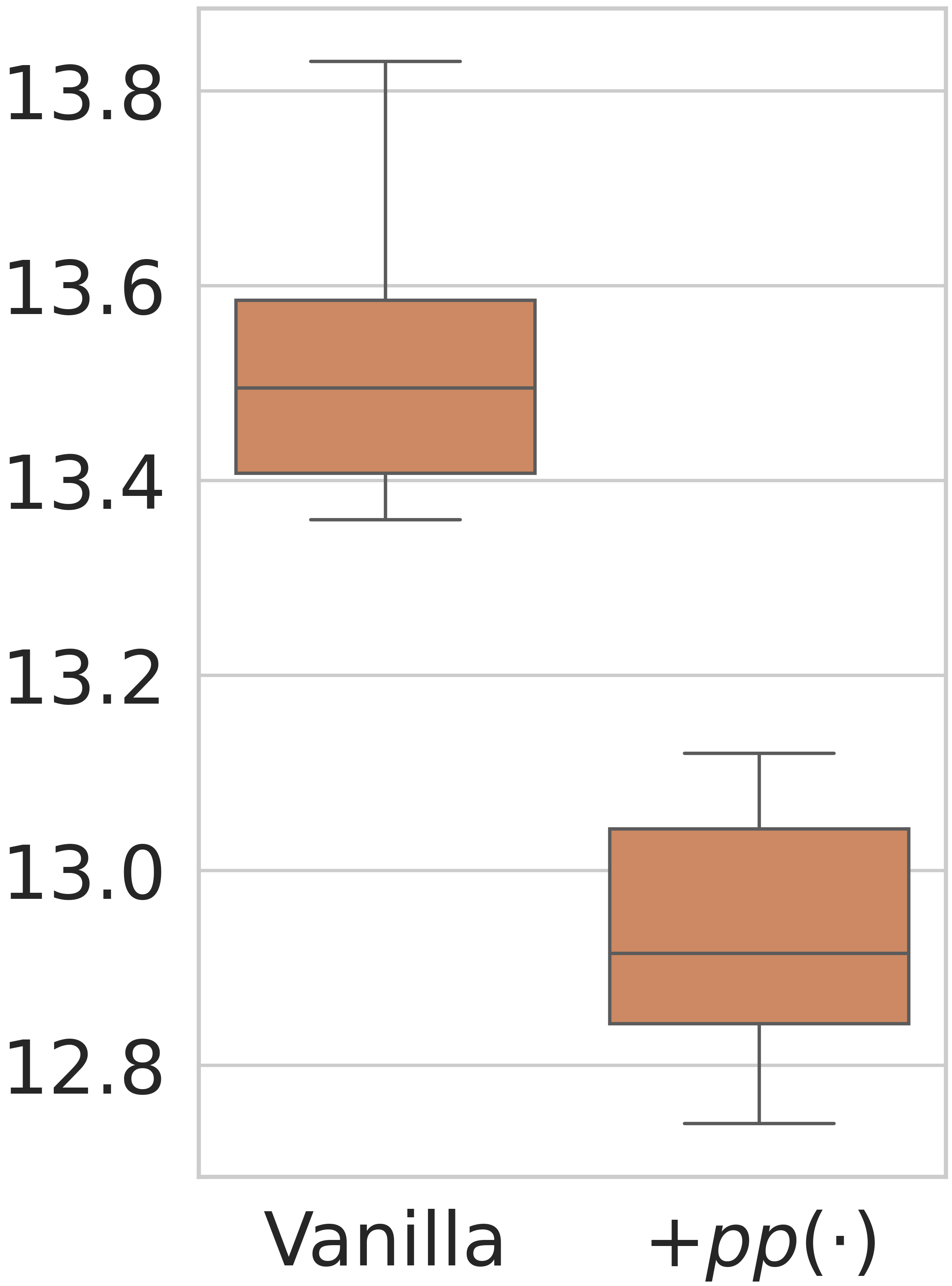}
        \caption{AMI}
        \label{fig:subfig3}
    \end{subfigure}

    % Second row
    \begin{subfigure}{0.1\textwidth}
        \centering
        \includegraphics[height=2.75cm]{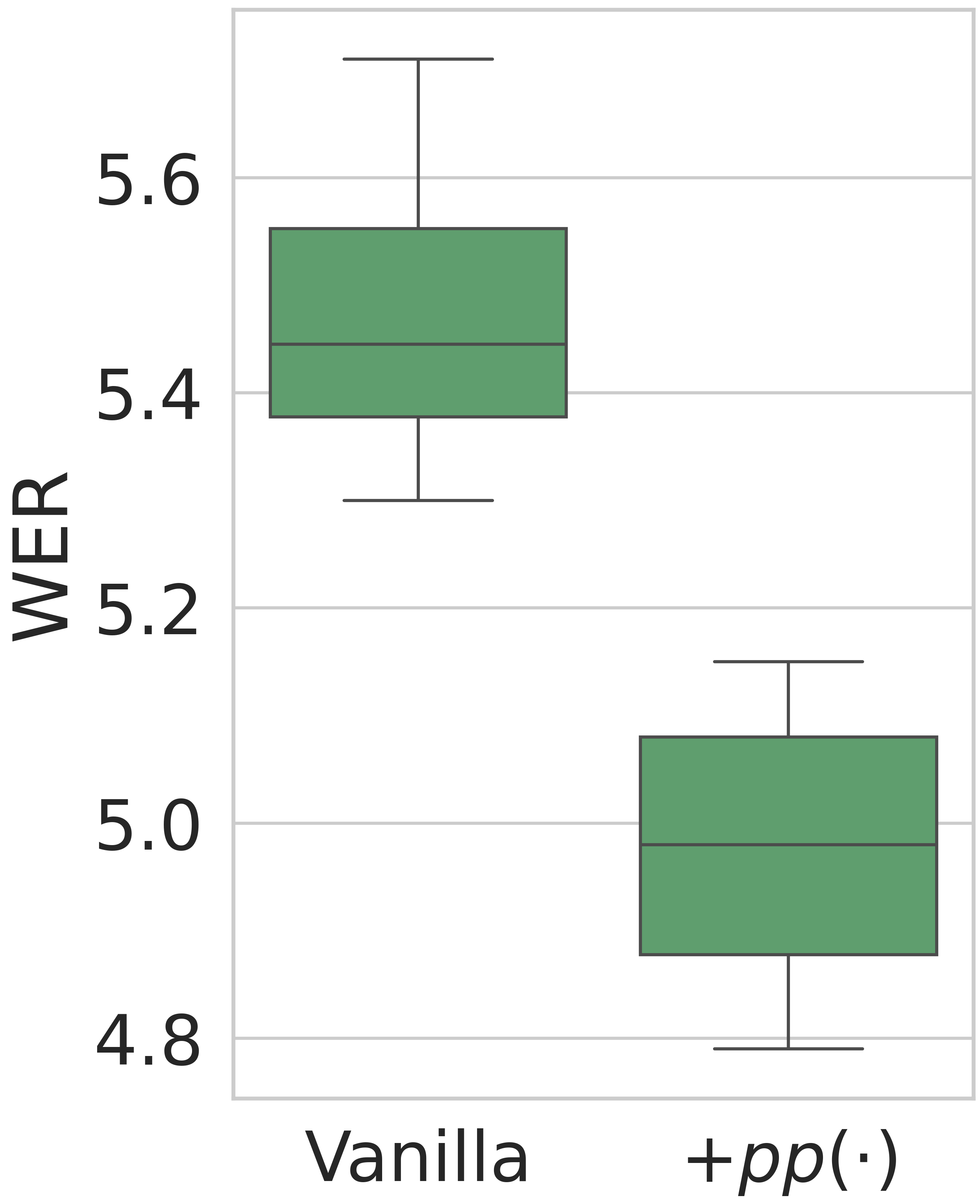}
        \caption{LS-O}
        \label{fig:subfig4}
    \end{subfigure}
    \hspace{2cm}
    \begin{subfigure}{0.1\textwidth}
        \centering
        \includegraphics[height=2.75cm]{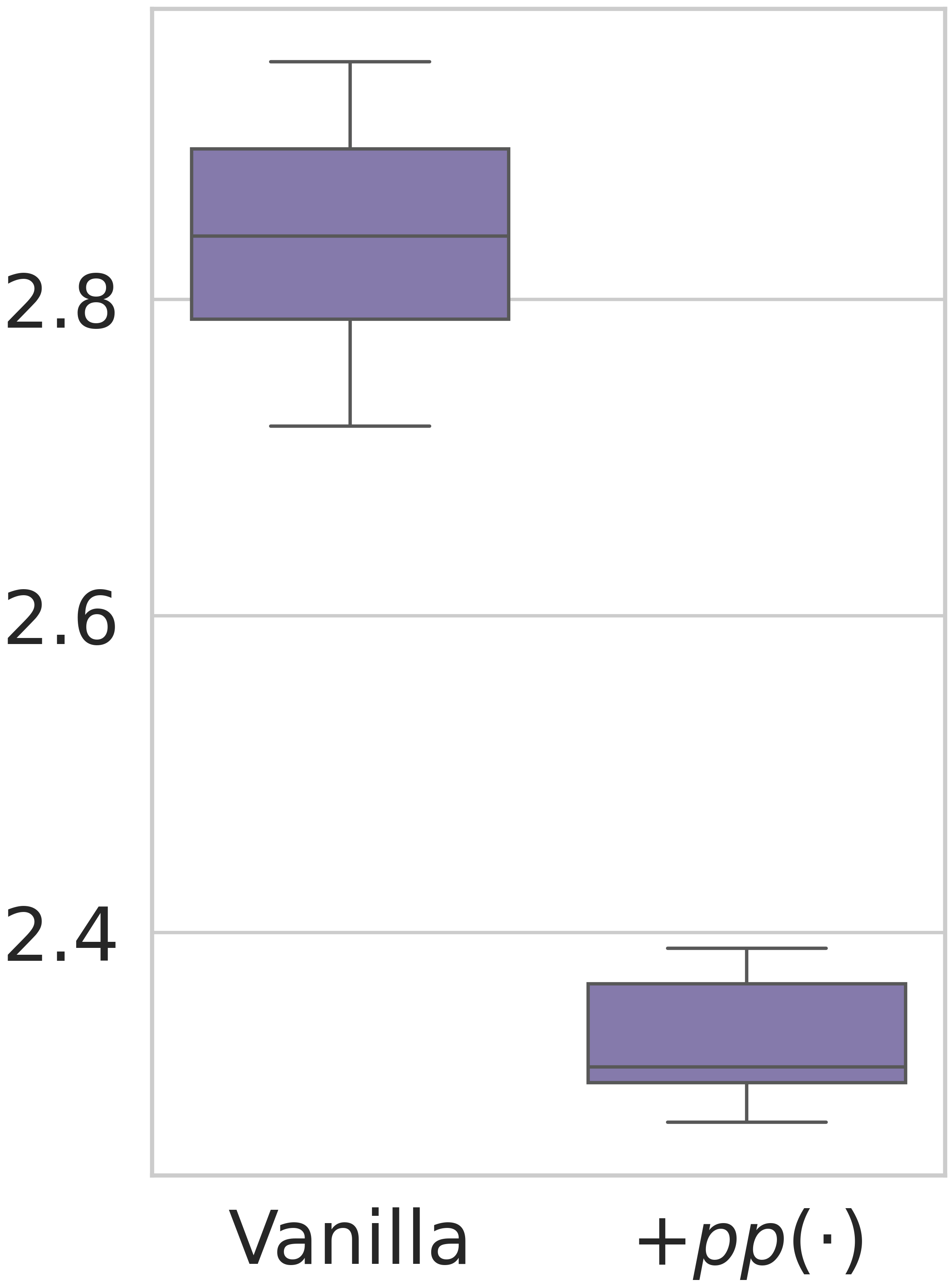}
        \caption{LS-C}
        \label{fig:subfig5}
    \end{subfigure}

    % \caption{Boxplots illustrating the impact of applying \textcolor{prompt}{$pp(\cdot)$} across different datasets. Each subplot compares WER (in \%) distributions among the different prompts before (\textit{vanilla}) and after applying \textcolor{prompt}{$pp(\cdot)$}.}
    \caption{Boxplots illustrating the impact of applying \textcolor{prompt}{$pp(\cdot)$} across different datasets. Each subplot compares WER (in \%) distributions among the different prompts before (\textit{vanilla}) and after applying \textcolor{prompt}{$pp(\cdot)$}. Improvements are statistically significant across all datasets according to paired statistical tests ($p<0.05$).\protect\footnotemark}

    \label{fig:mainfig}
\end{figure}

Incorporating \textcolor{prompt}{$pp(\cdot)$} reduces the gap between \textit{base} and \textit{best} prompts, producing WERs competitive with recent works. Crucially, \textcolor{prompt}{$pp(\cdot)$} mitigates the impact of suboptimal prompts: even the worst-performing prompt outperforms the best manual prompt alone (e.g., \textit{base}+\textcolor{prompt}{$pp(\cdot)$} vs. \textit{best}).
Figure~\ref{fig:mainfig} visualizes the before-and-after effect of using the prompt projector across the datasets, showing that \textcolor{prompt}{$pp(\cdot)$} consistently stabilizes and enhances performance regardless of the prompt.

%Overall, obtained results show that the new module is able to effectively learn a projection for all original prompt embeddings to improve the performance and mitigate their potential negative impact. %This simple extension is sufficient to reduce sensitivity to prompt choice and enhance the robustness of LLM-based ASR across diverse speech conditions.
Overall, our results show that the new module is able to learn prompt projections that improve performance for all original prompt embeddings, while mitigating the impact of unlucky prompt choices.

\begin{table}[t!]
    \centering
    %\scriptsize
    \footnotesize
    \caption{Main results. For reference, we include published results from recent LLM-based ASR systems. 
\textbf{Bold} indicates the best values in each group while \underline{\textbf{underlines}} marks the global best across our \textit{vanilla} and +\textcolor{prompt}{$pp(\cdot)$} groups.
}
    \renewcommand{\arraystretch}{1.0}
    \setlength{\tabcolsep}{6.8pt}
    \begin{tabular}{llcccccc}
        \toprule
        \multirow{2}{*}{\textbf{Method}} & \multirow{2}{*}{\textbf{Prompt}} & \multicolumn{5}{c}{\textbf{Word Error Rate (WER (in \%) $\downarrow$)}} \\
        \cmidrule(lr){3-7}
        & & CC & CH & AMI & LS-C & LS-O \\
        \midrule
        % \multicolumn{2}{l}{SLAM-ASR\cite{ma2024embarrassingly}}&-&-&- & 2.37 & 4.90 \\
        \multicolumn{2}{l}{SLM\cite{wang2023slm}}&-&-&15.14 & 2.60 & 5.00 \\
        \multicolumn{2}{l}{Q-Former\cite{10445874}}&-&-&- & 2.28 & 5.20 \\
        \multicolumn{2}{l}{Qwen-Audio\cite{chu2024qwen2}} &-&-&- & \textbf{2.04} & \textbf{4.20} \\
        \multicolumn{2}{l}{SpeechVerse\cite{das2024speechverse}} &-&-&- & 2.10 & 4.40 \\
        \multicolumn{2}{l}{SALMONN\cite{tangsalmonn}} &-&-&- & 2.10 & 4.90 \\
        
        \midrule
        \midrule
        \multirow{5}{*}{\rotatebox[origin=c]{90}{\textbf{\shortstack{vanilla}}}}
        & \textit{empty} & 12.75 & 27.00 & 13.88 & 2.84 & 5.40 \\
        & \textit{base} & 13.00 & 29.26 & 13.86 & 3.09 & 5.85  \\
        & \,\, \textit{+LoRA} & \textit{11.60} & \textit{28.18} & \textit{13.25} & \textit{2.46} & \textit{5.21} \\
        & \textit{best} & \textbf{11.81} & \textbf{25.26} & \textbf{13.36} & \textbf{2.72} & \textbf{5.30} \\
        % & \textit{best} & 12.00 & 25.56 & 2.75 & 5.38 & 13.42 \\
        & \,\, \textit{+LoRA} & \textit{11.43} & \textit{24.74} & \textit{12.79} & \textit{2.44} & \textit{4.95} \\
        \midrule
        \multirow{2}{*}{\rotatebox[origin=c]{90}{\textbf{\shortstack{+\\\textcolor{prompt}{$pp(\cdot)$}}}}} 
        & \textit{base} & 11.23 & 26.52 & 13.42 & 2.34 & 4.98  \\
        % & \textit{base} & 11.53 & 27.15 & 13.39 & 2.34 & 4.98  \\
        & \,\, \textit{+LoRA} & \textit{11.16} & \textit{25.86} & \textit{13.38} & \textit{2.33} & \textit{4.78} \\
        & \textit{best} & \textbf{11.23} & \textbf{24.73} & \textbf{12.74} & \textbf{2.28} & \textbf{4.79}  \\
        & \,\, \textit{+LoRA} & \textbf{\underline{11.14}} & \textbf{\underline{24.48}} & \textbf{\underline{12.72}} & \textbf{\underline{2.16}} & \textbf{\underline{4.66}} \\
        %& \textit{best-prm+LoRA} & & & & & \\
        \bottomrule
    \end{tabular}
    \label{tab:main_results}
\end{table}

\footnotetext{CC: $p=0.00135$; CH: $p=0.0221$; AMI: $p=1\times10^{-5}$; LS-O: $p=2.74\times10^{-4}$; LS-C: $p=3\times10^{-6}$; all with $p<\alpha=0.05$.}

\section{Conclusions}
\label{sec:conclusions}

In this work, we took a first step toward understanding and improving prompt robustness in LLM-based ASR. Through a systematic evaluation of manual prompts across multiple datasets, we showed that prompt choice has a large impact on transcription quality, with even minor wording or placement changes leading to notable differences in WER. This variability highlights the limitations of relying on fixed, hand-crafted prompts in practical systems.  

To address this, we introduced the \textit{prompt projector}, a lightweight, architecture-consistent extension that learns a common projection applied to manual prompt embeddings. Our experiments demonstrate that this simple design improves robustness to prompt choice without modifying the underlying system or introducing special prompt-engineering parameters or tokens.

%This work represents an initial exploration, and several unexplored questions remain. A natural next step is to compare the prompt projector with soft-prompt learning methods~\cite{prompt2023}, which add task-specific learnable tokens to the original prompt rather than learning a single projection. However, a fair comparison would require extensive experimentation not only across different hyperparameter settings (e.g., number of learnable tokens, initialization, placement), but also across multiple prompts and datasets, as done in this work. Such a study lies beyond the scope of this paper but would be valuable in a dedicated follow-up. Future work should also examine the applicability of the approach to different LLMs and speech encoders, as well as downstream tasks beyond transcription. Finally, an open question is whether a single projector could generalize, and to what extent, across prompts, tasks, or languages.
This work is an initial exploration, and several open questions remain. A natural next step is to compare the prompt projector with soft-prompt learning methods~\cite{prompt2023} that add task-specific learnable tokens to the original prompt, rather than learn a single projection. However, a fair comparison would require extensive experimentation across different hyperparameter settings (e.g., number of learnable tokens, initialization, placement), as well as multiple prompts and datasets, as done in this work. Such a study lies beyond the scope of this paper but would be valuable in a dedicated follow-up. Future work should examine the applicability of the approach to different LLMs and speech encoders, as well as to downstream tasks beyond transcription. Finally, an open question is whether a single projector can generalize across prompts, tasks, or languages.

%\section{Acknowledgments}
\noindent
\textbf{Acknowledgments} 
This work was supported by an Idiap Research Institute and Uniphore collaboration project. Part of this work was also supported by EU Horizon 2020 project ELOQUENCE. %(grant number 101070558).

%\vfill
%\pagebreak

\balance
\footnotesize
\bibliographystyle{IEEEbib}
\bibliography{main}

\end{document}